\documentstyle[prl,aps,amsfonts,amssymb,twocolumn,epsfig]{revtex}

\preprint{
\hfill  TH00.1
} 

\title{
Intermediate spin approach to the tunneling of nano-magnets.
} 

\author{S. E. Barnes }

\address{ Department of Physics, University of Miami, Coral Gables, 
Florida 33124 }

\begin{document}
\draft

\twocolumn[
\hsize\textwidth\columnwidth\hsize\csname @twocolumnfalse\endcsname

\date{\today} \maketitle
\begin{abstract}
The theory for the quantum tunneling of nano-magnets is developed 
within the intermediate spin framework.  Periodic magnetic effects are 
seen to reflect that associated with a change of flux by a single flux 
quantum $\Phi_{0}$.  Essential are Schr\"odinger cat wave functions 
which involve superpositions of different magnitudes of the applied 
magnetic field.  For systems in which the tunneling paths are not 
co-planar the theory leads to essentially complex magnetic fields 
although the expectation values of the fields remain real.  In general 
the ground states correspond to minimal uncertainty squeezed states. 
The degree of squeezing depends on the anisotropy parameters.
\end{abstract}


\vskip1.0pc]

\section{Introduction}

The quantum properties of very small magnets often called nano-magnets 
have generated a good deal of interest over the past decade.  Much of 
the activity has focused upon the demonstration of quantum mechanical 
tunneling in its various aspects\cite{1}.  Here attention is drawn to 
another purely quantum phenomenon namely {\it intermediate 
spin}\cite{khare,barnes}.  Intermediate spin comes as part of the 
baggage of intermediate statistics\cite{khare}, i.e., {\it anyons\/} 
and two dimensions.  The intermediate statistics {\it and spin\/} 
reflect a statistical parameter $\alpha_{s}$ which is conventionally 
$\alpha_{s}=0$ for bosons and whole integer spin and $\alpha_{s}=1$ 
for fermions and half-integer spin.  It does not seem to be generally 
appreciated that there are a good number of (real and numerical) 
experimental manifestations of intermediate spin in relatively simple 
magnet systems in a magnetic field.  The principal purpose of this 
paper is develop a formalism for quantum tunneling of nano-magnets 
based upon the theory of intermediate spin.

It will be show that a magnetic field directed along a suitable 
symmetry axis can cause, e.g., a physical whole integer nano-magnet to 
have an low energy level structure which is {\it identical\/} to the 
equivalent half-integer system, this including a Kramer's degeneracy.  
Zero and a certain\break
\begin{figure}[b!]
\vskip -10pt
\centerline{\epsfig{file=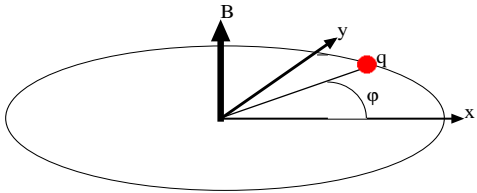,height=1.in,width=2.5in} } 
\vskip 10pt
\caption[toto]{A particle of charge $q$ is confined to a unit circle. 
A tunneling problem might be defined by adding 
a potential $V$ such that $V(\phi) = V(\phi + \pi)$. If this 
potential has minima at $\phi = 0$ and $\pi$ and barrier between is 
higher than the zero point energy then there will be a ground doublet 
with a splitting $\Delta$. A perpendicular field exerts no force on 
the particle but does change the slitting as explained in the text.
}
\label{f1}
\end{figure}
\noindent fields lying between the half-integer values correspond to 
whole integer points.  Intermediate fields correspond to intermediate 
spin.

The suppression of the ground doublet tunnel splitting which 
necessarily occurs at half-integer points has been previously 
recognized by Garg\cite{6}.  What has been labeled ``topological 
quenching'' is in fact just a pseudo-Kramer's degeneracy of the ground 
state.  The implications of intermediate spin are much stronger since 
{\it all\/} low lying levels must be degenerate {\it at exactly the 
same field}.

It is a commonplace observation, for the situation illustrated in 
Fig.~(1), that magnetic effects are periodic with period $\Phi_{0}$, 
the flux quantum.  For the problem explained in the figure caption in 
particular the tunnel splitting,
\begin{equation} 
\Delta = \Delta_{0} |\cos 2\pi \Phi/\Phi_{0}|.  
\label{un}
\end{equation}
For the magnetic systems the symmetry which will be considered in most 
detail here is that appropriate\cite{Barra96,21} to Fe${}_{8}$, 
Fig.~(2a).  This corresponds to the most general \break
\begin{figure}[b!]
\centerline{\epsfig{file=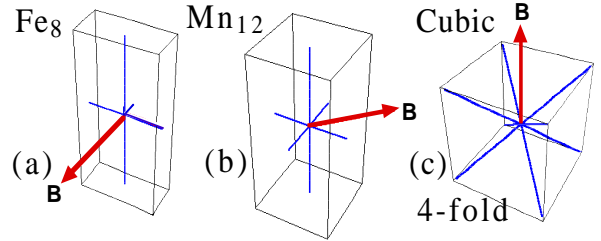,height=1.3in,width=3.1in} } 
\caption{The length of the axis indicates the energy gain when the 
spin $\vec S$ lies along that direction, thus (a) for the symmetry 
appropriate to Fe${}_{8}$ when all three axes are inequivalent the 
vertical $y$-axis is easy while the field is directed along the 
hardest $x$-direction.  For (b) appropriate to Mn${}_{12}$ with 
equivalent $x$ and $y$ directions, the $z$-direction is again easy.  
The field is along the hardest direction in the $x-y$-plane.  For 
cubic symmetry, illustrated is the possibility with the field along a 
hard four fold axis.  The perpendicular axes directed to the corners 
of the cube are all equivalent easy directions (in the absence of a 
field).  Intermediate spin effects also occur if the field is directed 
along a three fold axis.  }
\label{f2}
\end{figure}
\noindent   quadratic nano-magnet 
Hamiltonian:
\begin{equation}
{\cal H} = - D {S_{z}}^{2} 
 + E \left[{S_{x}}^{2} - {S_{y}}^{2}\right] 
\label{deux}
\end{equation}
which without loss of generality has two positive anisotropy 
parameters $E<D$, see the Appendix.  A magnetic field is directed 
along the hardest, i.e., $x$-direction. This is fully equivalent to the 
particle problem of Fig.~(1) with $\Phi/\Phi_{0} = g \mu_{B} 
H/4\hbar^{2}\sqrt{2E(D+E)}$.  The agreement with Eqn.  (\ref{un}) is 
illustrated by Fig.~(3).  The deviations from the $|\cos|$ behavior 
for larger fields is due to the finite value of $S$.  It is larger for 
the value $S=10$ relevant for Fe${}_{8}$.

Under an adiabatic change of magnetic field by $\Phi_{0}$ the $\hat 
S_{z}$ quantum number $m \to m+1$ for the angular momentum quantized 
along {\it the magnetic field direction}.  Whatever the mixture of 
states with different $m$, the period measured in terms of the 
magnetization along the field direction is automatically $\hbar g 
\mu_{B}$.  {\it However\/} it is a trivial consequence of time 
reversal invariance that the states with fluxes $\pm \Phi$ are 
degenerate and so there is the possibility to construct ``Schr\"odinger 
cat'' states which comprise a linear combination of states with the 
macroscopic field $\vec H$ in both its positive and negative 
directions.  In this case the magnetization period need no longer be 
$\hbar g \mu_{B}$.  It will be shown, with this symmetry, such a 
combination exists in all energy eigenstates and that the period is 
indeed a fraction of $\hbar g \mu_{B}$.

An even more bizarre phenomenon occur when either the field is not 
directed exactly along the appropriate symmetry axis and/or if the 
least action tunneling paths are not co-planar.  For these cases the 
energy eigenstates correspond to suitably weighted ``Schr\"odinger 
cat'' combinations with oppositely directed {\it complex\/} magnetic 
fields.  The complex weighting is such that the expectation value of 
the physical magnetic field remains real.  Thus, again for the 
Fe${}_{8}$ symmetry, a field $H_{t}$ which is perpendicular to both 
the hard and easy directions corresponds to a pure imaginary field $i 
H_{t}$ directed along the symmetry axis, i.e., the tunnel splitting is 
given by Eqn.  (\ref{un}) with $\Phi/\Phi_{0} = i g \mu_{B} 
H_{t}/4\hbar^{2}\sqrt{2E(D+E)}$ so that $\cos \to \cosh$.  In the 
presence of both fields\break
\begin{figure}[t!]
\centerline{\epsfig{file=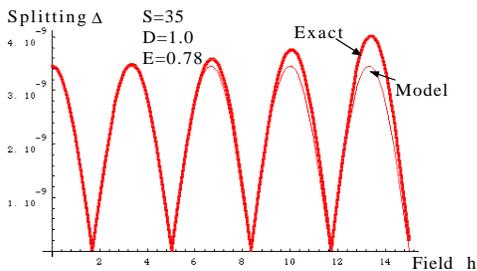,height=1.4in,width=2.5in} } 
\vspace{10pt} 
\caption[toto]{Properties are periodic in the field 
$h=H/g\mu_{B}\hbar$ for $S=35$.  The deviations from $\Delta = \Delta_{0} 
|\cos 2\pi \Phi/\Phi_{0}|$ become smaller as $S$ increases.  }
\label{f3}
\end{figure}
\begin{figure}[t!]
\vglue -1.0in
\centerline{\epsfig{file=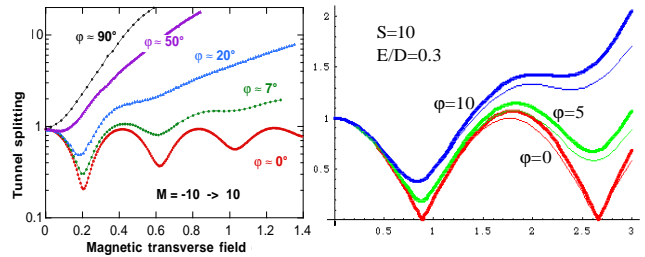,height=2.3in,width=3.5in} } 
\caption[toto]{To the right the thick line is approximation Eqn.~(1) 
with a complex field $B_{c} = B(\cos\theta + i\sin\theta)$.  The 
thinner line corresponds to exact diagonalization.  To the left is 
experiment\cite{21} with a logarithmic scale.  The imaginary part 
of the flux causes the tunnel splitting to increase with field.  }
\label{f4}
\end{figure}
\noindent $\Phi/\Phi_{0} = g \mu_{B} (H+i 
H_{t})/4\hbar^{2}\sqrt{2E(D+E)}$.  The resulting approximation is 
compared with the exact result and experiment \cite{21} in Fig.~(4).

For the much studied\cite{Sessoli93,thomas} molecular magnet 
Mn${}_{12}$ it is impossible to {\it not\/} have a complex field.  With 
the physical field in the direction indicated in Fig.~(2b) the 
resulting complex field has equal real and imaginary parts in the 
large $S$ limit, see Fig.~(5).  There are two equivalent tunneling 
planes which both make an angle $\theta = \pi/4$ with the applied 
field (and $\pi/2$ with each other).  The wave function for the planes 
is added so that there are two sets of paths, one from each plane, 
which are equivalent and which interfere.  The resulting expression for 
the tunnel splitting is shown in the figure caption.  As the figure 
illustrates this splitting increases very rapidly with field and by 
several orders for the $S$ oscillations.

The last important principle is illustrated by applying a field in the 
hardest direction for a system with cubic symmetry, Fig.~(1c).  This 
is a four fold axis and group theory insists that the lowest tunnel 
split multiplet comprises four states.  For whole integer $S$ it turns 
out that there are two singlets and one doublet, {\it however\/} for a 
half-integer point this must reduce to two doublets, reflecting the 
pseudo-Kramer's theorem.  That this is the case for large $S$ is
illustrated by the exact results of Fig.~(6).

Classically for magnet such as Fe${}_{8}$ the equilibrium\break
\begin{figure}[t!]
\centerline{\epsfig{file=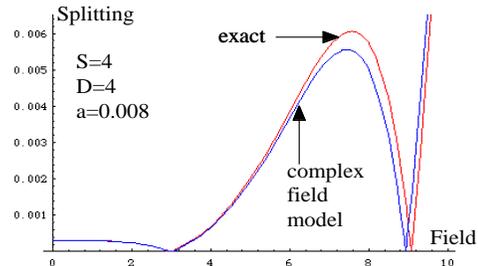,height=1.4in,width=2.5in} } 
\caption[toto]{The complex field result is $\Delta_{0}|\cosh(\pi h 
/p)\cos(\pi h /p)|$ where $ p \approx 2 S^{1/2} (12 a D^{3})^{1/4}$.  This 
corresponds to a complex magnetic field in which the real and 
imaginary parts are equal.  This exact result is indicated.  The 
difference becomes smaller with increasing $S$.}
\label{f5}
\end{figure}
\begin{figure}[t!]
\centerline{\epsfig{file=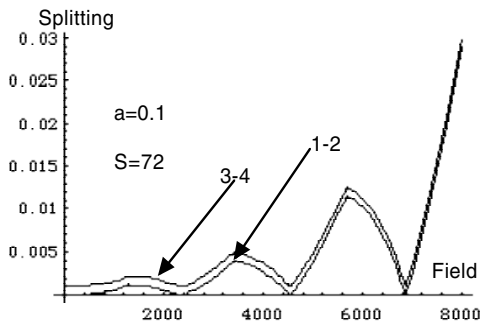,height=1.7in,width=2.5in} } 
\caption[toto]{For the four fold cubic axis there are four tunnel 
split levels.  Shown are the splittings between levels 1 and 2 and 
between 3 and 4.  The latter have been moved up by 0.001.  Illustrated 
is the fact that both splittings are zero for the same field.  This 
reflects a pseudo-Karmer's theorem for half-integer points.  (Notice 
that the splitting increases corresponding to a complex magnetic 
field.)}
\label{f6}
\end{figure}
\noindent magnetization lies along its easy axis and is maximal.  The 
existence of quantum fluctuations is reflected in the fact that the 
expectation value of this easy axis magnetization is $\hbar S$ and not 
$\hbar\sqrt{S(S+1)}$.  In the absence of a hard direction in the plane 
perpendicular to the easy axis, the quantum fluctuations measured by 
$\langle {S_{x}}^{2} \rangle$ and $\langle {S_{x}}^{2} \rangle \sim 
\hbar^{2} S/\sqrt{2}$ are equal and in fact {\it minimal}.  In the 
terminology used for fluctuations in quantum optics, these are minimal 
un-squeezed states.  Adding an anisotropy energy $E$, which creates a 
hard axis in this plane as is appropriate to Fe${}_{8}$, {\it 
squeezes\/} the quantum fluctuations so that $\langle {S_{x}}^{2} 
\rangle \ne \langle {S_{y}}^{2} \rangle$ but leaves minimal states.  
The present approach thereby lends itself to a natural definition of 
squeezed spin states\cite{Kitagawa}.

Apart from the simple periodicity\cite{21} of Eqn.~(\ref{un}) the intermediate 
spin effects described above have yet to be experimentally verified.

\section{Intermediate spin}

Returning to the basic charged particle problem of Fig.~(1), the 
relevant group is SO(2) which is equivalent to U(1), i.e., is abelian 
with a single operator $S_{z}$, the generator of rotations 
$e^{iS_{z}\phi}$, where the $z$-axis is perpendicular to the circle 
and where $\phi$ is the angle to the $x$-axis.  For a unit circle 
$S_{z} = p_{\phi}$ the momentum conjugate to $\phi$, i.e., $[\phi, 
p_{\phi}]=i\hbar$.  The eigenstates of $S_{z}$ and $p_{\phi}$ are 
$\psi_{m}(\phi) = {1 \over (2\pi)^{1/2}}e^{im\phi}$ but for particles 
on a circle there is no reason it insist on $m$ being an integer 
reflecting the fact that the spectrum of $S_{z}$ for $SO(2)$ is 
continuous.  In general for this {\it singular gauge} $\psi_{m}(\phi)$ 
is multi-valued.  A convenient single valued function can be defined 
by taking some convention for $\psi_{m}(\phi)$ in the interval 
$\{0,2\pi\}$ and then analytically continuing this function to the 
interval $\{-\infty,+\infty\}$.  For any physical potential $V(\phi) = 
V(\phi+2\pi n)$; $n=0, \pm 1, \pm 2 \ldots$, and implies that only the 
basis states $\psi_{{\alpha_{s}\over 2}+n}(\phi)$; $n=0, \pm 1, \pm 2 
\ldots$, mix.  It follows for a given statistical parameter 
$\alpha_{s}$ the energy spectrum is discrete.  This can be stated 
differently: Since $V(\phi)$ is periodic the solutions obey Floquet's 
(Bloch's) theorem, i.e., $\psi_{k}(\phi) = e^{ik\phi}u_{k}(\phi)$ 
where $u_{k}(\phi)$ is periodic.  The reciprocal lattice vector $K = 
{2\pi/ (a= 2\pi)} = 1$ and the energy $\epsilon(k)=\epsilon(k+nK); \ \ 
n=0, \pm 1, \pm 2 \ldots$ i.e., is periodic.  An examination of the 
Fourier components immediately implies $k= \alpha_{s}/2$.  Given that 
$V(\phi)$ has a point with reflection symmetry the $\psi_{\pm 
k}(\phi)$ are degenerate and the most general solution for a given 
energy is $A \psi_{k}(\phi) + B \psi_{-k}(\phi)$ for which the 
spectrum of $m = n \pm {\alpha_{s}\over 2}$, i.e., there are twice the 
number of observable $S_{z}$ values {\it except\/} importantly for the 
whole and half-integer points when $\alpha_{s} = 0$ or $1$.  This more 
general method of extrapolating between whole and half-integer spin 
which will be found applicable here.

In the {\it non-singular gauge\/} the basis set $\psi_{n}(\phi)$ has 
$n = \pm 1, \pm 2 \ldots$ and the charged particle encircles a 
magnetic flux $\Phi = \alpha_{s} \Phi_{0}/2$ where $\Phi_{0}$ is the 
appropriate flux quantum.  For the same $V(\phi)$ there exists an 
eigenstate $u_{k}(\phi)$ with the same energy as in the singular 
gauge, i.e., the energy spectrum are identical and two problems can 
mapped to each other and both reflect intermediate spin.  In the 
non-singular gauge the combination $A u_{k}(\phi) + B u_{-k}(\phi)$ 
comprises a linear combination of states with the flux $\Phi$ in both 
its positive and negative senses.  Evidently the expectation value of 
the flux $\langle \Phi \rangle$ is in general smaller than 
$\alpha_{s} \Phi_{0}/2$.  This possibility of a linear combination of 
different flux states does not seem to have been envisaged in 
connection with intermediate statistics although it would arise if, 
e.g., quantum coherence in SQUIDS could be observed.  It is realized 
in the systems, e.g., Fe$_{8}$, discussed here.

\section{Spin models}

It has become usual to re-write Eqn.~(2) with the $x$ and $z$-axes 
interchanged.  Thus with fields the general quadratic nano-magnet 
Hamiltonian becomes,
\begin{eqnarray}
&{\cal H} = - (D-E)& {S_{x}}^{2} + 2 E {S_{z}}^{2} \nonumber \\
&&- E S^{2}- h S_{z} - h_{\ell} S_{x}-h_{t}S_{y},
\label{una}
\end{eqnarray} 
see the Appendix.  Unless stated otherwise it will be assumed that the 
{\it physical\/} spin value $S$ is whole integer.  The value $S$ is 
that which would be assigned by the usual arguments, e.g., for a set 
of ferromagnetically ordered spins this is simply the sum of the spin 
values.  The field $h$ is parallel to the hardest and transverse to 
the easiest axis.  It is equivalent to that introduced in the 
non-singular gauge.  The other transverse $h_{t}$ and the longitudinal 
$h_{\ell}$ fields simply modify the potential $V$.

Since the length of $\vec S$ is fixed, classically the energy,
\begin{eqnarray}
&E(\theta, \phi) 
=
\hbar^{2} S^{2} \sin^{2}\theta \left[ - (D-E) \cos ^{2} \phi
+  2 E  \sin^{2} \phi \right]
\nonumber \\
&- h \hbar S \cos  \theta 
- h_{\ell} \hbar S \sin \theta \cos \phi
- h_{t} \hbar S \sin \theta \sin \phi
- E\hbar^{2} S^{2},
\label{deuxa}
\end{eqnarray}
is a function of the usual two angles $\theta$ and $\phi$, see Fig.~(7).

Quantum mechanically Eqn.  (\ref{deuxa}) is the energy expectation 
value for the coherent state $|\theta, \phi \rangle$ obtained by 
rotating $|S_{z} = S\rangle$ through $\theta, \phi$.  It is these 
coherent states which appear in the usual functional integral 
formulation\cite{1}.  {\it However\/} the $|\theta, \phi \rangle$ are 
{\it very\/} much over-complete.  In fact the states $| \phi \rangle 
\equiv |\theta={\pi \over 2} , \phi \rangle = (2S+1)^{1/2} 
\sum_{m=-S}^{S} e^{-im\phi}|m \rangle $, where $|m \rangle \equiv 
|S_{z} = m \rangle$, remain over-complete.  If $ | \psi \rangle = 
\sum_{m=-S}^{S} a(m) |m \rangle$ then $\psi(\phi) \equiv \langle \phi 
| \psi \rangle = \sum_{m=-S}^{S} e^{im\phi} a(m)$, i.e., is a Fourier 
transform of the $m$-space wave function $a(m)$.  The function 
$\psi(\phi)$ is defined on the interval $\{-\infty,+\infty\}$ and the 
inverse transform $\psi(m) \equiv {1\over 2\pi} 
\int_{-\infty}^{+\infty}\, d\phi \psi(\phi) = \sum_{m^{\prime}} 
\delta(m - m^{\prime}) a(m^{\prime})$ so that $ | \psi \rangle = 
\int_{-\infty}^{+\infty}\, dm\, \psi(m) |m \rangle$, i.e., $\psi(m)$ is 
a definition of the $m$-space wave function appropriate when $m$ is 
considered as a continuous variable.  Clearly $\hat S_{z} \psi(\phi) = 
\int_{-\infty}^{+\infty}\, dm e^{im\phi} \hbar m \psi(m) = p_{\phi} 
\psi(\phi) $ where $ p_{\phi} = - i \hbar {d \over d \phi}$ is the 
same as that defined earlier.  Also evident is that multiplication by 
$e^{i\phi}$ amounts to a displacement of $m$ by unity and so, e.g., 
$\hat S^{+} = e^{i\phi}[\hbar^{2}S(S+1) - p_{\phi}(p_{\phi}+1)]^{1/2} $.

\section{Topological considerations}

In terms of the unit sphere, the system of coordinates $\phi, 
p_{\phi}$ might be considered as a particular projection in which the 
axis of quantization plays the role of the North-South axis as shown 
in Fig.~(7).  In these terms, the angle $\phi$ is evidently the 
longitude while $p_{\phi}$ is the projection on the $z$-axis and is 
therefore the sine of the latitude.  That these are canonical 
coordinates and momentum emphasizes the evident fact that the unit 
sphere is the phase rather than simply the coordinate space. 

In order to calculate the Berry phase\cite{Berry} a unit vector $\vec 
R$ is directed along the easy, i.e., $x$-axis for Eqn.~(\ref{una}) 
(the $z$-axis for Eqn.~(\ref{un})).  It is particularly easy to 
determine this phase for an adiabatic process in which the system is 
rotated about the $z$-axis (the $x$-axis for Eqn.~(\ref{un})).  For 
the singular gauge, if the wave function comprises a single Floquet 
(Bloch) wave $ e^{ik\phi}u_{k}(\phi)$ this is trivially
\begin{equation}
\Theta_{B} = 
2\pi k = 2\pi {\alpha_{s}\over 2}
	\label{berry}
\end{equation}
which is the obvious generalization of the result $\Omega S$ where 
$\Omega$, here $2\pi$, is the solid angle subtended by the path of $\vec 
R$.  {\it However\/} for the combination $A \psi_{k}(\phi) + B 
\psi_{-k}(\phi)$ the result is $\tan \Theta_{B} = [(A-B)/(A+B)] \tan 
2\pi {\alpha_{s}\over 2}$ which agrees with $\Omega {\alpha_{s}\over 
2}$ only at integer and half-integer points.

\section{Pseudo Kramer's operators}

If it {\it is\/} possible to transmute a whole into a half-integer 
spin then at the half-integer points there should exist an\break
\begin{figure}[b]
\centerline{\epsfig{file=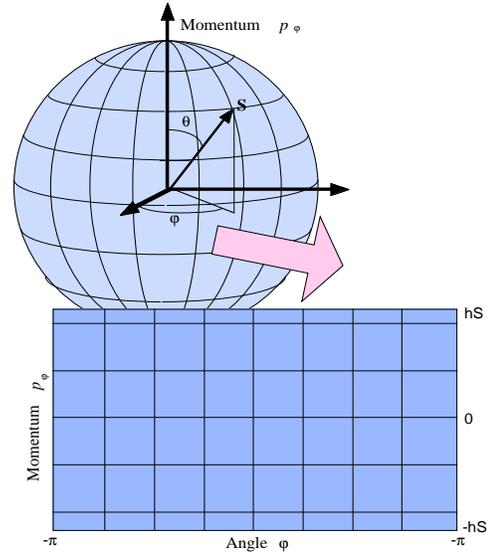,height=3in,width=2.5in} } 
\caption[toto]{Mapping of the unit sphere to a two dimensional 
rectangle
}
\label{f7}
\end{figure}
\noindent operator $K$, the equivalent of the time reversal operator 
for $h=0$.  This must commute with $\cal H$ and if $|E_{n} \rangle$ is 
an eigenstate then $K|E_{n} \rangle$ should be orthogonal (or zero).  
For the current model Preda and Barnes\cite{Preda00} have constructed 
such operators.

The simplest case is with $h_{\ell} = h_{t} = 0$ and $h = 
\hbar \sqrt{2E(D+E)}$. The operator $K_{1} = Tr$ where $T$ is the 
conventional time reversal operator and where 
$$
r = \sqrt{D+E} S_{x} + i\sqrt{2E} S_{y}.
$$
It is easy to show that with this field the negative semi-definite,
$$
{\cal H} = - (Tr)^{2}
$$
so that trivially $[{\cal H}, K_{1}]=0$ {\it independent of $S$ and the 
parameters $D$ and $E$}.  As will be shown in Sec.  VII, with $h_{\ell} 
= h_{t} = 0$ there is an eigenstate of the form $|E_{n} \rangle \equiv 
\sum_{p} a(2p) |2p \rangle$, where $S_{z} |m \rangle = \hbar m |m 
\rangle$, i.e., only states with even values of $m$ are present.  Since $r$ 
only has matrix elements between even and odd such states it follows 
that $K_{1} |E_{n} \rangle$ is orthogonal to $|E_{n} \rangle$ and 
hence for this value of the field all states are double degenerate 
{\it or\/} $K_{1} |E_{n} \rangle =0$.  Clearly this latter state has 
energy $E_{n} = 0$.  It is evident that when $E=0$ there is only one 
such state and by a continuity argument this must be the case for all 
$E$.  It follows that there are $S$ degenerate pairs when $h = 
\sqrt{2E(D+E)}$ fully mimicking a half-integer spin except for the 
highest energy singlet.

It is therefore a remarkable fact that this degeneracy for fields 
which are {\it exact\/} multiples of $\hbar \sqrt{2E(D+E)}$ exists for 
all but the highest energy states for arbitrary values of integer $S$ 
and independent of the existence of a tunneling barrier, etc.  For 
example it is possible to explicitly construct the $S=1$ matrices and 
show that for $h = \hbar \sqrt{2E(D+E)}$ the ground state is 
degenerate independent of the parameters $E$ and $D$.

It is to be concluded that this degeneracy corresponds to a {\it 
hidden symmetry\/} of this class of Hamiltonian, akin to time reversal 
symmetry for the half-integer equivalent in zero field.  This 
pseudo-time-reversal symmetry then implies a pseudo-Kramer's 
degeneracy.  

\section{Quantum fluctuations - Squeezed states}

If $S$ are not too small then $\vec S$ remains always close to the 
classical energy minima at $\theta = \pi/2$ and $\phi = 0$ or $\pi$, 
and to an excellent approximation near the first of these classical 
ground states Eqn.  (\ref{deux}), after dropping a constant 
$E^{0}=-\hbar^{2} S^{2}(D-E)$ and with $h_{t}=h_{\ell}=0$, simplifies to,
\begin{equation}
{\cal H}
=
 (D+E) {p_{\phi}}^{2}  + h p_{\phi} +
 \hbar^{2} (D-E) S^{2} \phi^{2}.
\label{trois}
\end{equation}
The field can be absorbed into the kinetic energy so
\begin{eqnarray}
&{\cal H}
=
 (D+E)\left( {p_{\phi}+\delta }\right)^{2} 
+& \hbar^{2} (D-E) S^{2} \phi^{2}\nonumber \\
&& - (D+E){{\delta}^{2}},
\label{troisa}
\end{eqnarray}
where $\delta = h/2(D+E)$.  Recall, $\delta$ corresponds to {\it 
displacement\/} of $\delta $ in $m$-space, however it is not equal to 
$\alpha_{s}/2$.  In $\phi$ space it adds a phase factor $e^{i\delta 
\phi }$ to the wave function.  Even for finite $h$, this is just a 
harmonic oscillator with a ground state energy
\begin{equation}
\epsilon_{n} = E^{0} - {h^{2} \over  4(D+E)} + 
(n+{1\over2})\omega_{0},
\label{energy}
\end{equation}
where $\omega_{0} = S \hbar \sqrt{D^{2}-E^{2}}$ and where only the 
last term is of quantum origin.  The wave function
\begin{equation}
a(\phi) 
=
\left({S \over \pi} \right)^{1/2}\left({D-E \over D+E}
\right)^{1/4} e^{i\delta \phi} 
e^{-S \left({D-E \over D+E}\right)^{1/2} {\phi^{2} \over 2}}.
\label{quatre}
\end{equation}
Directly, the magnetic moment, 
\begin{equation}
\langle S_{z} \rangle = 
\langle p_{\phi} \rangle = \hbar \delta = \hbar {h\over 2(D+E)}.
\end{equation}
The same result follows from differentiating $\epsilon_{n}(h)$.

This ground state wave function of a harmonic oscillator necessarily 
corresponds to the minimum of the uncertainty relationship $\Delta 
\phi \Delta p_{\phi} \ge \hbar/2$.  This permits a natural definition 
of {\it squeezed\/} spin states.  The transverse spin fluctuations are 
a minimum, e.g., for $E=0$, $\sigma_{x} = \sigma_{y} = \hbar 
(S/2)^{1/2}$ (e.g., ${\sigma_{x}}^{2} \equiv \langle {S_{x}}^{2} \rangle - 
\langle {S_{x}} \rangle^{2}$).  The $\vec S$ can be pictured as having 
fluctuating transverse components $\sim \hbar (S/2)^{1/2}$ such that 
the total magnitude square is correctly $\hbar^{2}[S^{2}+ 2\times 
(S/2)] = \hbar^{2} S(S+1)$.  Clearly the two quadrature components 
$\sigma_{x}$ and $\sigma_{y}$ have equal uncertainties and this 
minimum uncertainty state is {\it un-squeezed}.  For finite $E$, 
$$
{\sigma_{ x}}^{2} = \left({\hbar^{2} S \over 2}\right)\sqrt{D-E \over 
D+E}; \ \ \  {\sigma_{y}}^{2} = \left({\hbar^{2} S \over 
2}\right)\sqrt{D+E \over D-E}.
$$
Maximum squeezing occurs as $E \to D$. 

\section{Formulation with the $|S_{z}\rangle$ basis}

The zero point energy associated with the transverse fluctuations is 
$\omega_{0}/2$ and correctly the low lying excited states have 
energies $(n+(1/2))\hbar\omega_{0} \sim \hbar^{2} SD$ above the ground 
state.  However from Eqn.  (\ref{deuxa}) the energy barrier is also 
$\sim S^{2}\hbar^{2} D$ and this would apparently correspond to a 
problem with a tunneling amplitude of order $e^{-S/2}$ which is not 
inconsistent with the Fourier transform of Eqn.~(\ref{quatre}).  {\it 
However\/} the unsqueezed states when $E=0$ are in fact eigenstates of 
$S_{x}$, i.e., in this case tunneling is absent.

It is necessary to carefully formulate the problem in terms of the 
basis set $|m\rangle$, i.e., in $m$-space.  If $|\psi \rangle = 
\sum_{m=-S}^{+S}a(m)|m\rangle$, Schr\"odinger's equation in this 
basis can be written as ($h_{\ell} = h_{t} =0$):
\begin{eqnarray}
&\left(\epsilon - (2E\hbar^{2} m^{2} - h\hbar m) \right) a(m) \nonumber \\
&= 
{1\over 4}\hbar^{2} (D-E)\Big[
M_{m}^{m+1} M_{m+1}^{m+2} a(m+2) \nonumber \\
&+
M_{m}^{m-1} M_{m-1}^{m-2} a(m-2)\nonumber \\
&+[{M_{m}^{m+1}}^{2}+{M_{m}^{m-1}}^{2}]a(m)
\Big], 
\label{chain}
\end{eqnarray}
where $M_{s}^{t} = \left[ S(S+1) - st \right]^{1/2}$.  The structure 
of this equation is worth noting.  It has the form of {\it two\/} 
tight binding models, one for even $m$ values and the other for odd 
$m$.  The two chains are not connected as illustrated\break
\begin{figure}[t!]
\centerline{\epsfig{file=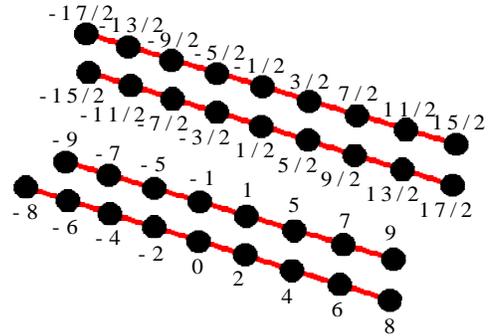,height=1.8in,width=2.9in} } 
\caption[toto]{Schr\"odinger's equation Eqn.~(\ref{chain}) reflects 
two disconnected chains. For whole integer $S$ one chain corresponds 
to even and the other odd sites and are inequivalent. In constrast for 
half-integer $S$ the chains have $n \to -n$ symmetry. That these 
chains have the same eigenenergies reflects Kramer's theorem.
}
\label{f8}
\end{figure}
\noindent in Fig.~(8).  
This is not a periodic solid since (i) the diagonal energy increases 
steady going away from $m=0$ (ii) the chains have ends at $m = \pm S$.  
This is rather a pair of discrete harmonic oscillator problems for 
large $S$ when (ii) is not important since the wave function is well 
localized near $m=0$.

Making a continuum approximation, e.g., $a(m\pm 2) \approx a(m) \pm 2 
(da/dn) + 2 (d^{2}a/dn^{2})$, valid if $S^{2}(D-E) \gg 2E$, and 
assuming large spin $S \gg 1$, this reduces to
\begin{equation}
\Big( 2E\hbar^{2} {m}^2 - \epsilon \Big) a(m) 
+ 4 \hbar^{2} S^{2}(D-E){d^{2} a(m) \over d m^{2}} =0.
\end{equation}
This describes a simple harmonic oscillator with wave function $a(m) = 
e^{-m^2Q/2S}$ where, $Q = \sqrt{(D+E)/(D-E)}$ characterizes the degree 
of squeezing.  This should be recognized as the Fourier transform of 
Eqn.~(\ref{quatre}).  There is evidently no tunneling barrier and no 
tunnel ground doublet splitting!  {\it However\/} there {\it is\/} a 
degeneracy since this same equation results for both the odd and even 
site chains.  The splitting of these two states disappeared upon 
making the continuum approximation.  As will become evident below, the 
discrete problem described by Eqn.~(\ref{chain}) has asymptotic 
corrections which differ as the $m$ arguments are even of odd.  It is an 
interesting observation that the discrete harmonic oscillator and 
tunneling problems are equivalent.

The wave function for the eigenstate $|S_{x} = S\rangle$ is $a^{0}(m) 
= {1\over 2^{S}} \left[{(2S)!  \over (S-m)!(S+m)!}\right]^{1/2}$ and 
$a^{0}(m)$ has to be a solution to Eqn.~(\ref{chain}) with $h=E=0$, this 
corresponding to the absence of tunneling.  Using Stirling's 
approximation this reduces to $a(m) \propto e^{-m^2Q/2S}$ with $E=0$, 
however the deviations from this approximation are very important.  
The absence of tunneling when $h=E=0$ is put in evidence by a 
transformation: $a(m) = a^{0}(m) f(m)$.  After some algebra 
Schr\"odiner's equation for $f(m)$ reduces to:
\begin{eqnarray}
	 & \left(\epsilon - (2E\hbar^{2} m^{2} - h\hbar m +E^{0})\right)   f(m)
	  =
{1\over 4} \hbar^{2} (D-E)
	\nonumber  \\
	 &\times \Big[
(S(S-1)+m^{2})  (f(m+2)+f(m-2)-2f(m))
	\nonumber  \\
	 & -
n(2S-1)
(f(m+2)-f(m-2)) \Big].
	\label{trans}
\end{eqnarray}
When $h=E=0$ this evidently admits the solutions $f(m) = 1$ and $f(m) = 
e^{i\pi m}$ corresponding to the degenerate states $|S_{x} = \pm S\rangle$. 

\section{Intermediate spin formulation}

With again the definition $f(\phi) = \sum_{m} e^{im\phi} f(m)$ this 
yields the {\it exact\/} Schr\"odinger's equation:
\begin{equation}
	\Big({\cal E} - 2E{p_{\phi}}^{2}\Big) 
f(\phi) =  V(\phi, p_{\phi}) f(\phi)
	\label{cinq}
\end{equation}
with
\begin{eqnarray}
&V(\phi, p_{\phi})	 = {1\over 2} (D-E) \times\nonumber \\
&
\Big[
2\sin^{2}\phi(\hbar^{2} S(S-1)+{p_{\phi}}^{2})
-
i \hbar (2S-1)\sin 2\phi p_{\phi}  
\Big],
	\label{six}
\end{eqnarray}
where ${\cal E} = (\epsilon - E^{0})$.  This is of the intermediate 
spin form but with a momentum dependent potential, $V(\phi, 
p_{\phi})$.

The effect of the transverse field $h$ is twofold (i) near the 
classical ground state the spin sees a longitudinal field $S(D-E)$ 
along with the transverse $h$ and this tilts the axis of quantization 
towards the $z$-direction.  (ii) The field $h$ modifies the tunneling 
amplitude.

In accounting for (i) the term $2E{p_{\phi}}^{2}$ is not relevant and 
so what is required is a solution of,
\begin{equation}
({\cal E} + hp_{\phi})f(\phi) = V(\phi, p_{\phi}) 
f(\phi).
	\label{sept}
\end{equation}
Needed is a small rotation about $y$-axis.  The new wave function is 
defined by $f(\phi) = (1-(\alpha_{s} p_{\phi}/2\hbar S)) g(\phi)$ 
where $h = \hbar (D-E) \alpha_{s}/2$, this admixes, e.g., a little of 
the first excited into ground state.  For large $S$, $g(\phi)$ obeys
\begin{equation}
{\cal E}g(\phi) = V(\phi, p_{\phi}+\hbar {\alpha_{s} \over 2}) 
g(\phi),
\label{huit}
\end{equation}
i.e., the field has been replaced by the displacement ${\alpha_{s} \over 
2}$. 

In order to account for (ii), a similar displacement is included in the 
full equation. If  $g(\phi)$ is the solution of 
\begin{eqnarray}
	&\Big({\cal E} - 2E\left( p_{\phi}+\hbar {\alpha_{s} \over 2}\right)^{2}
&+\hbar^{2}  E {{\alpha_{s}}^{2} \over 2} \Big) 
g(\phi)	\nonumber \\
&& =  V(\phi, p_{\phi}+\hbar {\alpha_{s} \over 2}) 
g(\phi),
	\label{neuf}
\end{eqnarray}
then $f(\phi) = (1-(\alpha_{s} p_{\phi}/2\hbar S)) g(\phi)$ is the solution 
to 
\begin{eqnarray}
	&\Big({\cal E} - 2E{ p_{\phi}}^{2} + &2 \hbar (D+E) {\alpha_{s} \over 2} p_{\phi}
 \Big) 
f(\phi)	 \nonumber \\
&&=  V(\phi, p_{\phi}) 
f(\phi).
	\label{dix}
\end{eqnarray}
Following the discussion of intermediate spin, the shift determines 
the wave vector $k = {\alpha_{s}/2}$.  In turn the statistical 
parameter must be chosen to give ground state energy determined 
without tunneling, i.e., Eqn.~(\ref{energy}).  Required is that 
$\hbar^{2} E {{\alpha_{s}}^{2} \over 2} = {h^{2} \over 4(D+E)}$, so 
$k= {\alpha_{s} \over 2} =h/2\hbar \sqrt{2E(D+E)}$.  This fixes the 
Zeeman term in Eqn.  (\ref{dix}); $2 \hbar (D+E) {\alpha_{s} \over 2} 
p_{\phi} = \sqrt{(D+E)/2E} h p_{\phi}$.  Since, in general, this is 
larger than $h p_{\phi}$ the solution in the non-singular gauge $A 
(1-(\alpha_{s} p_{\phi}/2\hbar S)) u_{k}(\phi) + B (1+(\alpha_{s} 
p_{\phi}/2\hbar S)) u_{-k}(\phi)$ involves a linear combination of both bare 
fields $\pm \sqrt{(D+E)/2E} h$ such that the expectation value is 
equal the physical applied field.  This requires $A = 1 + 
\sqrt{2E/(D+E)}$ and $B = 1 - \sqrt{2E/(D+E)}$.  The total energy for 
level $n$ and wave vector $k$ is,
\begin{equation}
E_{n}(k) = E^{0} - {h^{2} \over 4(D+E)}
+ \left(n+{1 \over 2}\right)\hbar \omega_{0} + \epsilon_{n}(k).
\label{onze}
\end{equation}
The first two terms are purely classical in origin while the third 
corresponds to the harmonic motion about the bottom of the well.  It 
is the last term which reflects the tunneling between wells.

The problem therefore reduces to the determination of 
$\epsilon_{n}(k)$ and when the tunneling approximation is well 
justified this reflects the tight binding approximation.  Actually for 
the case consider so far, with only a transverse field $h$, the 
problem has a higher symmetry, see Fig.~(9).  The potential $V(\phi, 
p_{\phi})$ has a $\phi$ period of $a=\pi$ rather than $2\pi$ so the 
reciprocal space unit vector $K=2$.  Since $K$ corresponds to the 
smallest relevant displacement in $m$-space, it follows that a given 
solution only involves $m$ values which differ by $2$, as already 
noted in Sec.  VII. The tunneling energy,
\begin{equation}
\epsilon_{n}(k) = {\Delta_{0} \over 2} \cos  \pi k
\label{douze}
\end{equation}
where $\Delta_{0}/2$ is the yet undetermined (diagonal) matrix element 
for tunneling for level $n$.  The evaluation of this will be addressed 
in Sec.  XIII. There are {\it two\/} solutions which are acceptable for 
$h=0$, namely $k = {\alpha_{s} \over 2} = 0$ corresponding to even $m$ 
values {\it and\/} $k = {\alpha_{s} \over 2} =1$ and odd $m$.  Clearly 
a shift by one brings the set of integers $m$-values back to itself.  
These two $k$ values correspond to $\epsilon(k) = \pm {\Delta_{0} 
\over 2}$ and so the zero field tunnel splitting is reflected by the 
parameter $\Delta_{0}$.  For finite transverse fields $h$ the 
corresponding $k = {\alpha_{s} \over 2} = (h/2\hbar \sqrt{2E(D+E)})$ and $k = 
1 + {\alpha_{s} \over 2} = 1+(h/2\hbar \sqrt{2E(D+E)})$ and so the tunnel 
splitting is,
\begin{equation}
\Delta = \Delta_{0} \cos  { \pi h \over 2 \hbar \sqrt{2E(D+E)}}
\label{trieze}
\end{equation}
equivalent to a result first obtained by Garg\cite{6} and attributed 
to topological quenching.

\section{Half-integer spin}

Nowhere in the development has use been made of the assumption that 
$S$ is an whole integer.  The only difference when $S$ is a 
half-integer is that the values\break
\begin{figure}[t!]
\centerline{\epsfig{file=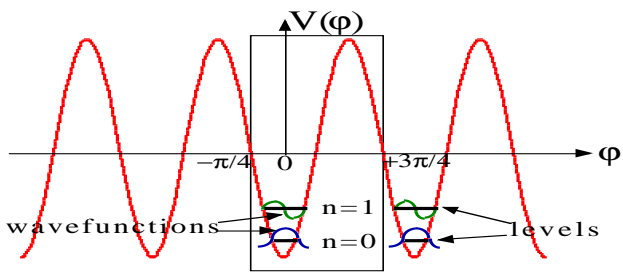,height=1.4in,width=3.3in} } 
\vskip 5pt
\caption[toto]{When $h_{t} = h_{\ell}=0$ the symmetry is higher and 
the period is $\pi$ rather than $2\pi$.  The states with label 
$n=0,1,\ldots$ near the bottom of the wells are to a good 
approximation those of a harmonic oscillator.  The box contains the 
unit cell, however in the singular gauge it is useful to continue the 
wave function and potential into the entire interval 
$\{-\infty,+\infty\}$.  
}
\label{f9}
\end{figure}
\noindent of $m$ are also half-integer and thus 
the solutions correspond to $k = {1\over 2} + {\alpha_{s} \over 2} = 
{1\over 2} + (h/2 \hbar \sqrt{2E(D+E)})$ and $k = - {1\over 2} + {\alpha_{s} 
\over 2} = - {1\over 2} + (h/2\hbar \sqrt{2E(D+E)})$ which yields a splitting,
\begin{equation}
\Delta = \Delta_{0} \sin  { \pi h \over 2\hbar \sqrt{2E(D+E)}}
\label{quatorze}
\end{equation}
i.e., the cosine is replaced by a sine.

\section{Continuation of solution}

The solution described above is based on a small $h$ approximation it 
is valid in the very large $S$ limit.  That the matrix elements of 
$h_{p} = p h_{1}; \ h_{1} = 2\hbar \sqrt{2E(D+E)}$ to excited states be 
small requires $2E p < S^{2}(D-E)$.  For the first non-trivial whole 
integer point when $h = h_{1}$ this is equivalent to the requirement 
that the continuum approximation be valid, or that the shifts $\pm 
{\alpha_{s}\over 2}$, or $\pm k$, for a field $h=h_{p}$ are smaller 
than the width of Gaussian wave function $e^{-m^{2}Q/2S}$.  The wave 
function $A (1-(\alpha_{s} p_{\phi}/2\hbar S)) u_{k}(\phi) + B 
(1+(\alpha_{s} p_{\phi}/2\hbar S)) u_{-k}(\phi)$ is therefore a valid 
approximation for $h=h_{1}$ for not too small $S$.  However it will 
fail for modest values of $2E$ and large fields $h$.  The theory can 
be extended to cover such larger fields by a process similar to 
analytic continuation.  The current approximation can always be 
developed to construct the essentially exact solution $a^{1}(m)$ for 
$h=h_{1}$.  The dimensionless moment $m_{1} \approx h/2 \hbar (D+E)$ for 
$h_{1}$ can be determined.  Re-defined are the states $|\phi \rangle = 
\sum_{m} e^{i(m-m_{1})\phi} |m \rangle $ so that $S_{z} = -i \hbar 
(d/d\phi) + \hbar m_{1}$.  This translates back to the origin the wave 
function $a^{1}(m)$.  Using the original $a^{0}(n)$ an equation for $f(n)$ 
is developed.  Important is that in this equation there can be no term 
linear $n$ reflecting an effective field, since this would imply a 
finite displacement ${\alpha_{s} \over 2} = k$ and hence a finite 
first derivative of the tunneling energy.  This contradicts the 
assumption that the initial $h=h_{1}$ reflects a whole integer point 
and hence a maximum in the tunnel splitting.  The rather obvious point 
being made is that the whole (and half-) integer points can be 
represented without a mixture of different effective fields while in 
between these points this is not possible.  Continuing the development 
for points near to $h_{p}$, the new wave vector $k = {\alpha_{s}\over 
2} = \delta h /2\hbar \sqrt{2E(D+E)}$ where $\delta h = h - h_{p}$.  (The 
period cannot change since otherwise the current solution would 
predict a whole integer point which is not exactly at zero field.)  
Always the shifts are smaller than the Gaussian width and the new 
${\alpha_{s}\over 2}$ differs from the old value by an integer and is 
hence equivalent.  The fact that there is more than a single fashion 
by which to construct the wave function reflects the fact that each of 
the sets $|\phi \rangle$ independent of $m_{1}$ are (over-) complete.  
Important is the observation that away from the whole and half-integer 
points the wave function reflects an essential mixture of fields.  The 
description with a minimum of difference has fields $h_{p} \pm 
\sqrt{(D+E)/2E} \delta h ; \ \delta h = h - h_{p}$, and where $h_{p}$ 
corresponds to the closest whole or half-integer point.

\section{Effect of other field components}

Consider first the effect of $h_{\ell}$ alone.  This adds a potential,
\begin{equation}
V(\phi, p_{\phi})  \Rightarrow V(\phi, p_{\phi}) + \hbar S h_{\ell}  \cos \phi
\label{quinze}
\end{equation}
and removes the symmetry between $\phi = 0$ and $\pi$.  The period of 
the potential is now the full $a=2\pi$, with $K=1$, {\it but\/} with two 
wells per unit cell.  The harmonic levels near $\phi= \pi$ have 
quantum numbers designated by $n$ and are higher than those near $\phi 
= 0$ and which will be labeled $n^{\prime}$.  Ignoring tunneling 
these have energies,
\begin{equation}
E_{n} =E_{0} - {h^{2} \over 4(D+E)} + \hbar n h_{\ell} 
+ \left(n+{1 \over 2}\right)\hbar \omega_{0},
\label{seize}
\end{equation}
and,
\begin{equation} 
E_{n^{\prime}} = E_{0} - {h^{2} \over 4(D+E)} - \hbar n^{\prime} h_{\ell} 
+ \left(n^{\prime}+{1 \over 2}\right)\hbar \omega_{0}.
\label{dixsept}
\end{equation}
For a given $h$ there is only one solution with $k = {\alpha_{s} \over 
2} = (h/2 \hbar \sqrt{2E(D+E)})$, the solution displaced by unity 
being the same state.  There are now two possibilities (i) $E_{n} \ne 
E_{n^{\prime}}$ and there is a {\it very\/} narrow band formed about\break
\begin{figure}[t!]
\centerline{\epsfig{file=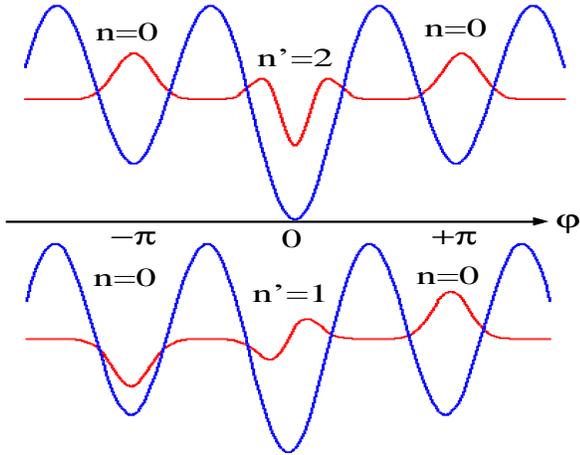,height=2.4in,width=3.1in} } 
\vspace{10pt} \caption[toto]{The effective potential $h_{t} =0$, 
$h_{\ell}\ne 0$ has two inequivalent minima per unit cell.  While the 
wells near $\pm \pi$ are physically equivalent, in the singular gauge 
the wave function is multi-valued and so the phase when $\phi$ 
increases by $2\pi$ corresponds to the Berry phase $2\pi k$ and is 
relevant.  When the field $h=0$ the wave functions can be made real.  
Illustrated are the two different resonant cases.  Above is the case 
when $n=0$ and $n^{\prime}=2$, i.e., when both wave functions are even 
and there is no change in sign.  In contrast, below $n=0$ and 
$n^{\prime}=1$, i.e., one wave function is odd and so the wave must 
change sign as its $\phi$ argument advances by $2\pi$.  }
\label{11}
\end{figure}
\noindent both these energies.  The energy differences have a {\it very\/} small 
dependence on $k$ and hence $h$, see below.  (ii) When $E_{n} \approx 
E_{n^{\prime}}$ there is resonant tunneling.  There are again two 
cases (iia) when both $n$ and $n^{\prime}$ are odd or even and (iib) 
when one of $n$ and $n^{\prime}$ is odd and the other even.

As illustrated in Fig.~(10) there are two tunneling matrix elements 
one $\Delta_{n,n^{\prime}}$ for the barrier between $\phi = 0$ and 
$\pi$ and $\Delta_{n^{\prime},n}$ for that between $\phi = -\pi$ and 
$0$ (or $\pi$ and $2\pi$).  With first $h=(h_{t} =) 0$, 
Schr\"odinger's equation admits real solutions, $\psi(m)$ or 
$\psi(\phi)$, and so these matrix elements can be made real.  For case 
(iia) when the symmetry of the wave functions is the same the 
tunneling matrix elements $\Delta_{n,n^{\prime}}= 
\Delta_{n^{\prime},n} \equiv \Delta_{0}$.  For finite $h$, for the 
Floquet (Bloch) wave function, the phase advances by $2\pi k$ from one 
cell to the next while to agree with the smaller unit cell appropriate 
when $h_{\ell}=0$ the phase should advance by $\pi k$ from one well to 
the next within the cell.  This leads to $\Delta_{n,n^{\prime}}= 
e^{i\pi k} \Delta_{0}$ and $ \Delta_{n^{\prime},n} = e^{-i\pi 
k}\Delta_{0}$.

At this point it is possible to introduce the second transverse field 
$h_{t}$ by slight of hand.  It is observed that if $E$ changes sign in 
the original version of $\cal H$, Eqn.  (\ref{deux}), then the two 
transverse axes change role and thus $h$ becomes the equivalent of 
$h_{t}$.  If $\cal H$ is considered to be a function of a complex 
parameter $E$ the solution might be analytically continued from 
positive to negative values.  In particular $k = (h/2\hbar 
\sqrt{2E(D+E)})$ becomes pure imaginary, i.e., $k \to ik_{t} = 
i(h_{t}/2\hbar \sqrt{2E(D-E)})$ and it follows that 
$\Delta_{n,n^{\prime}}= e^{i\pi (k+ik_{t})} \Delta_{0}$ with $ 
\Delta_{n^{\prime},n} = e^{-i\pi (k+ik_{t})}\Delta_{0}$.  This 
information can be represented by a two level model:
$$
{\cal H} = (E_{n}-E_{n^{\prime}}) S_{z} + {\Delta_{0}\over 2}
\left[ e^{i\pi (k+ik_{t})} S^{+} + 
e^{-i\pi (k+ik_{t})}S^{-}\right],
$$
where the center of gravity energy ${1\over2} 
(E_{n}+E_{n^{\prime}})$ has been dropped.  Degenerate 
perturbation theory then yields,
\begin{equation}
\epsilon_{k} = 
 \pm {1\over2}
\sqrt{(E_{n}-E_{n^{\prime}})^{2}+
4{\Delta_{0}}^{2}|\cos \pi (k+ik_{t})|^{2}}
\label{dixhuit}
\end{equation}
or at resonance $\epsilon_{k} = \pm {\Delta_{0}\over 
2}|\cos  \pi (k+ik_{t})|$, with $k = {h \over 2 \hbar 
\sqrt{2E(D+E)}}$ and $k_{t} = {h_{t} \over 2 \hbar \sqrt{2E(D-E)}}$.  
The $h_{\ell}= h_{t} =0$ result is recovered when $n = n^{\prime}$.  
That this is a good approximation is illustrated by comparison with 
exact results in Fig.~(4).  When for (iib) one wave function changes 
sign while the other does not it is necessarily the case, for 
$h=h_{t}=0$, that $\Delta_{,n^{\prime}}= - \Delta_{n^{\prime},n} 
\equiv \Delta_{0}$ whence,
\begin{equation}
\epsilon_{k} = 
 \pm {1\over2}
\sqrt{(E_{n}-E_{n^{\prime}})^{2}+
4{\Delta_{0}}^{2}|\sin \pi (k+ik_{t})|^{2}}
\label{dixneuf}
\end{equation}
and $\epsilon_{k} = \pm {\Delta_{0}\over 2}|\sin  \pi 
(k+ik_{t})|$ at resonance and, Eqn.~(\ref{quatorze}), this is 
equivalent an original half-integer $S$.  For case (i), Eqns.  
(\ref{dixhuit},\ref{dixneuf}) serve to illustrate that when 
$|E_{n}-E_{n^{\prime}}| > 2 \Delta_{0}\sin \pi k$, e.g., 
the upper sign of Eqn.  (\ref{dixhuit}) gives $\epsilon_{k} \approx 
(E_{n}-E_{n^{\prime}}) + 2({\Delta_{0}}^{2}\sin^{2} \pi 
k/(E_{n}-E_{n^{\prime}}))$ which contains a 
$k$-dependent correction which is smaller than those for cases (ii) by 
a factor of $\Delta_{0}/ (E_{n}-E_{n^{\prime}})$.  This 
factor is usually {\it extremely\/} small for the region between 
crossings.

\section{Combined intermediate spin effects}

Both $h$ and $h_{\ell}$ cause the effective spin $S-{\alpha_{s}\over 
2}$ to alternate between effective whole and half-integer spin and the 
two fields can be combined so that, e.g., a field $h = \hbar 
\sqrt{2E(D+E)}$ transmutes a whole into a half-integer spin.  Adding 
$h_{\ell} = \hbar \sqrt{D^{2}-E^{2}}$ brings say the $n=0$ level in 
the higher well into resonance with the $n=1$ of the lower well and 
the system again behaves as whole integer with a maximal tunnel 
splitting.  {\it However\/} the fashion in which the two fields work 
is different.  The transverse field $h$ leads to a smoothly changing 
(Berry) phase shift of $\delta = \pi (h \hbar / \sqrt{2E(D+E)}$ in the 
boundary condition $\psi(\phi + 2\pi) = e^{i\delta}\psi(\phi)$ so that 
values of $h$ which are not multiples of $\hbar \sqrt{2E(D+E)}$ correspond 
to an intermediate spin value.  In contrast the phase shift caused by 
$h_{\ell}$ results from the changes in sign of the wave function for 
the odd $n$ levels, i.e., when this field brings two levels into 
resonance $\delta = \pi (n - n^{\prime})$ which causes the effective 
spin value to alternate between whole and half-integer without a 
meaningful intermediate case.

\section{The tunnel splitting.}

There are three regimes for which different approximate methods must 
be used in order to calculate the tunnel splitting.  (i) If $E \ll D$, 
and for quantization along the easy axis, the maximal tunnel splitting 
$\Delta_{0}$ is calculated treating $E$ as a perturbation, see below.  
(ii) If $S^{2}(D-E) < 2E$, i.e., very close to the point $E=D$ the 
energy levels can be calculated using the hard axis for quantization 
and $(D-E){S_{y}}^{2}$ is a weak the perturbation.  To a very good 
approximation the energies are $2D\hbar^{2} m^{2}- (h^{2}/16E)$ where $m= 
n+(h/4\hbar E)$ and $n$ is an integer\cite{barnes}.  Thus, e.g., the energy 
between the ground state and first excited state for $h=0$ is $2D$, 
i.e., very large and nothing to do with tunneling.  As a function of 
field the splitting is a sawtooth.  Otherwise (iii) for $E \approx D$ 
the problem corresponds to a dominant hard axis anisotropy with again 
$(D-E){S_{y}}^{2}$ small.  This is the case considered in considerable 
detail by Preda and Barnes\cite{Preda99}.  Following a method 
described by S. Coleman\cite{25} it is possible to calculate the 
tunneling matrix elements near to any crossing $m,m^{\prime}$.  The 
result is\cite{Preda99},
\begin{eqnarray}
&{\Delta_{0}\over 4} = t_{m,m^{\prime}} \nonumber \\
& =\hbar \omega_0
{2^{n\over 2+1 }\over \sqrt{n!\pi}} 
\left( {2 S^2 K_{\perp}\over K_{\parallel}}\right)^{n+1}
\exp\left(
-2\sqrt2\sqrt{S^2 K_{\perp}\over K_{\parallel}}\right).
\label{pertun}
\end{eqnarray}

For the remaining case (i) the energy levels are $E_{m} = -D\hbar^{2} 
m^{2};\ m=\pm S, \pm(S-1), \ldots \pm 1, 0$.  There are two cases (a) 
when the two levels in resonance $n$ and $n^{\prime}$ both have the 
same parity and (b) when the parity is different.  The whole integer 
case (a) is simplest since $\Delta_{0}$ can be determined for $h=0$.  
The matrix element of $2E{S_{x}}^{2}$ between the states $m$ and $m+2$ 
is
\begin{equation}
	p_{m,m+2} = {E\over 2}\hbar^{2}  M_{m}^{m+1} M_{m+1}^{m+2},
	\label{matrix}
\end{equation}
recalling that $M_{m}^{m+1} = [S(S+1) - m(m+1)]^{1/2}$. Thus the expression 
for the tunnel splitting is,
\begin{eqnarray}
	 & \Delta_{0} = \left({\hbar^{2} E \over 2}\right)^{n^{\prime} - n\over 2}
M_{n}^{n+1} M_{n+1}^{n+2}{1 \over E_{n+2} - E_{n}}
\nonumber  \\
	&M_{n+2}^{n+3} M_{n+3}^{n+4}{1 \over E_{n+4}- E_{n}}
\ldots \nonumber  \\
 & M_{n^{\prime}-4}^{n^{\prime}-3} M_{n^{\prime}-3}^{n^{\prime}-2}
{1 \over E_{n^{\prime}-2}- E_{n}}\nonumber  \\
&M_{n^{\prime}-4}^{n^{\prime}-3} M_{n^{\prime}-3}^{n^{\prime}-2}
{1 \over E_{n^{\prime}-2}- E_{n}}
M_{n^{\prime}-2}^{n^{\prime}-1} M_{n^{\prime}-1}^{n^{\prime}}.
	\label{tunnel}
\end{eqnarray}
For a half-integer point, i.e., case (b) it is observed that, with $k 
= h/2\hbar \sqrt{2E(D+E)}$, Eqn.~(\ref{quatorze}) gives a splitting,
\begin{equation}
\delta (h) = \pi \Delta_{0} {h \over 2\hbar \sqrt{2E(D+E)}},
\label{fieldsplit}
\end{equation}
for small enough $h$.  This must agree with a perturbation theory in 
which $h$ can be taken to the smallest quantity.  With the current 
axis of quantization the term $hS_{x}$ implies a matrix element of 
$hM_{m}^{m+1}/2$ between the levels $m$ and $m+1$.  The field provides 
one such matrix element to the expression,
\begin{eqnarray}
&\delta(h) &= h \sum_{m=n+1}^{n^{\prime} -1}
\left({\hbar^{2} E \over 2}\right)^{n^{\prime} - n\over 2}
\nonumber \\
&& \ldots
M_{n}^{n+1} M_{n+1}^{n+2}{1 \over E_{n+2}- E_{n}}
M_{n+2}^{n+3}\nonumber M_{n+3}^{n+4}{1 \over E_{n+4}- E_{n}}
\ldots
\nonumber \\
&&\ldots 
M_{m-1}^{m-1}M_{m-1}^{m}{1 \over E_{m}- E_{n}}
{M_{m}^{m+1} \over 2} \nonumber \\
&&\times {1 \over E_{m+1}- E_{n}}
M_{m+1}^{m+2} M_{n+2}^{n+3}
\ldots \nonumber \\
&&\ldots M_{n^{\prime}-4}^{n^{\prime}-3} M_{n^{\prime}-3}^{n^{\prime}-2}
{1 \over E_{n^{\prime}-2}- E_{n}}
M_{n^{\prime}-2}^{n^{\prime}-1} M_{n^{\prime}-1}^{n^{\prime}}
\nonumber \\
&&
\equiv C(n,n^{\prime}) h  
\left({E \over 2(D+E)}\right)^{n^{\prime} - n-1\over 2},
\label{perttunnel}
\end{eqnarray}
so that,
\begin{eqnarray}
&\Delta_{0}& = C(n,n^{\prime}) {\hbar^{2} \sqrt{2E(D+E)} \over \pi} 
\left({E \over 2(D+E)}\right)^{n^{\prime} - n-1\over 2}\nonumber \\
&&= E {C(n,n^{\prime})  \over \pi} 
\left({E \over 2(D+E)}\right)^{n^{\prime} - n\over 2},
\label{delta}
\end{eqnarray}
which has the same behavior as Eqn.~(\ref{tunnel}) for transitions 
between the same parity.  In both cases,
\begin{equation}
\Delta_{0} \sim 2\hbar^{2} E e^{- |{n^{\prime} - n\over 2}| S_{0}};
\  \ \ \ 
S_{0} = \ln \left({E \over 2(D+E)}\right),
\label{split}
\end{equation}
i.e., $2E$ is the attempt frequency and the action agrees with that 
which is obtained for the ground state to ground state tunneling using 
the functional integral method\cite{1}.  There is little interest in 
reducing the expressions Eqns.~(\ref{tunnel}) and (\ref{delta}) since 
precise values for splitting are easily obtained numerically.  It is 
an interesting result that the attempt frequency is $2\hbar^{2} E$ and 
not perhaps the characteristic energy $2\hbar^{2} \sqrt{2E(D+E)}$ or 
$\hbar^{2} \sqrt{D^{2} - E^{2}}$.

\section{Other symmetries}

The issue that intermediate spin effects are manifested even when there 
is no evident easy plane has been raised in the Introduction.  
Consider cubic symmetry, the appropriate Hamiltonian,
\begin{equation}
{\cal H} = a \left[{S_{x}}^{2}+ {S_{y}}^{2}+ {S_{z}}^{2}\right]
+ h S_{z}
\label{cubic}
\end{equation}
has eight classical equilibrium directions, Fig.~(1c) which with $a$ 
positive {\it do not\/} coincide with the axes.  Quantum mechanically, 
in the absence of a field the system resonates between each of these 
equilibrium directions.  However in the language of Sec.~VII, there 
are only {\it four\/} chains and for large spin only four closely 
spaced levels.  Symmetry dictates that in zero field the chain with 
sites $\ldots, -3,1,5,\ldots$ has levels which are degenerate with the 
similar chain with $n \to -n$.  To a good approximation for large $S$ 
Schr\"odinger's equation is:
\begin{eqnarray}
	 &[\epsilon- (a\hbar^{4} m^{4} +{3a\hbar^{4} \over 8}(S(S+1)-m^{2}))] a(m)
	\nonumber  \\
	& = {a\hbar^{4} \over 16}\sum (S(S+1)-(m\pm 2)^{2}) a(m\pm 4)
	\label{schcubic}
\end{eqnarray}
where $a(m)$ is the wave function.  An analytic approach to the 
solution of this equation is facilitated by a modest field $h$ and 
large $S$.  This suppresses the tunneling between the state with 
positive and negative projections on the $z$-direction.  Now tunneling 
occurs between the four classical equilibrium directions with a 
negative $S_{z}$ component.  It is the four least action paths between 
these which map to a circle.  Making the continuum approximation near 
the minimum in the the diagonal energy with a negative such projection 
defines a discrete simple harmonic oscillator with a potential $\sim 
2a\hbar^{4} S^{2} (m-m_{0})^{2}$ where $m_{0}$ is the equilibrium 
dimensionless magnetization.  In Fourier space this becomes the 
kinetic energy and the potential $\sim a\hbar^{4} S^{4}\cos 4\phi$ so 
that there are four equivalent minima.  Apart from this reduction of 
the period by a factor of two, the problem is then fully equivalent to 
that considered in Sec.~VIII and the tunnel splitting quasi-periodic 
as illustrated in Fig.~(6).  As the field increases the minimum energy 
directions bend towards the $z$-axis and this reduces the potential 
barrier between the equivalent wells and increases the kinetic energy, 
hence the steady increase in the tunnel splitting.  The separation 
between the half-integer points when the splitting is zero is seen to 
be $\sim 4 \hbar^{4} S^{4} a$ but varies since the coefficient of the 
$m^{2}$ kinetic energy term also increases with field.  In fact there 
are two level crossings at small fields which occur before the large 
field approximation is valid and which are not visible in this figure.  
At least for larger fields these half-integer points really do 
correspond to a Kramer's degenerate half-integer spectrum for the low 
lying quartet as is explained in the Introduction and the Fig.~(6) 
caption.

The symmetry appropriate to Mn${}_{12}$ has an easy $z$-axis with 
equivalent $x$ and $y$ directions so that the 
Hamiltonian\cite{thomas},
\begin{equation}
	{\cal H} = - D {S_{z}}^{2} + a \left[{S_{x}}^{2}+ {S_{y}}^{2} \right]
	+ h S_{x},
	\label{Mn}
\end{equation}
where, with $a$ positive, the $x$-directed transverse field is along a 
hard direction in the $x-y$-plane.  Classically in zero field there 
are equivalent equilibrium directions which have the magnetization 
parallel or anti-parallel to the $z$-direction.  A finite $h$ does no 
alter the equivalence of these two directions, the magnetization 
simply gains a component opposite to the field direction.  However the 
field {\it does\/} have an effect on the tunneling paths.  In zero 
field there are four least action paths which pass via one of the easy 
directions in the $x-y$-plane.  With a field in this particular 
direction and $a$ positive there remain only two equivalent paths 
which pass via an easy direction for which the magnetization has a 
component in the direction away from the field.  It is these two least 
action paths which map to the intermediate spin circle.  Here the 
paths are not co-planar and the field $h$ will enter in the potential 
{\it but\/} without altering the symmetry.  

It is possible to analytically account for all four tunneling paths 
for large $S$ and not too large a $h$ by imagining that there are two 
crossed circles.  A given circle comprises two easy directions in the 
$x-y$-plane and $z$-axis, i.e., the two planes containing the circle 
make an angle of 45${}^{o}$ to the applied field.  The wave function 
for each circle is calculated and the two are added to give the final 
result.  Because the field is at 45${}^{o}$ to each circle the wave 
vector has equal real and imaginary parts and the net result is that 
displayed in the caption to Fig.~(5).

\section{Experimental verification of field mixing}

Certainly the most striking claim is that the wave function $A 
(1-(\alpha_{s} p_{\phi}/2\hbar S)) u_{k}(\phi) + B (1+(\alpha_{s} 
p_{\phi}/2\hbar S)) u_{-k}(\phi)$ for an intermediate spin system is a 
mixture of a state with the local applied field of magnitude $h_{p} 
\pm \sqrt{(D+E)/2E}\delta h; \ \delta h = h - h_{p}$, i.e., fields 
both larger and smaller than the distant applied field $h$.  (With an 
equivalent claim for $h_{t}$ which involves $\sqrt{(D-E)/2E} 
h_{t}$.)\break
\begin{figure}[t!]
\centerline{\epsfig{file=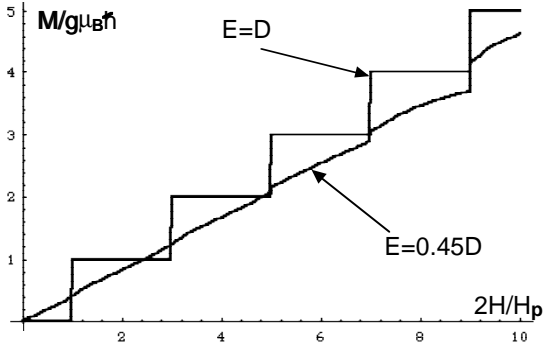,height=1.8in,width=2.9in} } 
\vspace{10pt} \caption[toto]{The dimensionless moment $m(h)$.  The 
field is divided by half the period so that all steps coincide and 
$S=5$.  The moment period is exactly unity for $E=D$ but otherwise is 
less and implies a Schr\"odinger's cat with a superposition of fields.  
}
\label{f11}
\end{figure}
\noindent  For a particle on a ring this would imply a distortion in the near 
field with a range of order of the ring diameter.  Unfortunately the 
spin has no natural dimension and the ring diameter must be scaled to 
zero.  {\it Experimentally\/} the existence of this superposition is 
easy to establish.  In the absence of the superposition the wave 
function would have a simple Floquet form: $e^{ik\phi}u_{k}(\phi)$, 
where $k={\alpha_{s} \over 2}$ is determined, modulo unity, from the 
experimental tunnel splitting: $\Delta_{0}|\cos \pi k|$.  Theory gives 
$k = {\alpha_{s} \over 2}= \delta h /2 \hbar \sqrt{2E(D+E)}$ which 
agrees with experiment\cite{21} for Fe$_{8}$.  Following the Fourier 
transform to $m$-space the factor $e^{ik\phi}$ corresponds to a 
displacement of $k$ and would lead to a dimensionless magnetic moment,
\begin{equation}
m_{\rm non-mix} = k = {\alpha_{s} \over 2} = {\delta h \over 
2\hbar \sqrt{2E(D+E)}},
\label{moment}
\end{equation}
again modulo unity.  In the limit of large $S$ the theory accounting 
for the superposition leads to
$$
m_{mix} \approx {h \over 2\hbar (D+E)} 
$$ 
which is smaller except when $E=D$. This has yet to be verified 
experimentally.

A more subtle way of stating the same principle goes as follows: The 
transverse field $h$ couples to the physical $S_{z}$, i.e., this 
operator in the non-singular gauge, and which has a spectrum with a 
period $\hbar$ or in units of magnetic moment $g \mu_{B}\hbar$ where 
$g$ is the g-factor.  Thus if $M_{z}$ is the measured moment in the 
direction of the transverse field, $m = M_{z}/g \mu_{B}\hbar$, the 
dimensionless magnetization, has a natural period of unity, a 
statement equivalent to the observation that physical effects are 
periodic in the flux quantum $\Phi_{0}$.  This would be the case if 
the state did not consist {\it precisely\/} of a superposition of 
field directions.  Because of the admixture, the measurable period is 
reduced to $ \Delta m \equiv m_{1} = \sqrt{2E \over D+E} \le 1$ as 
illustrated with exact results in Fig.~(11).  It should be noted that 
the value of $m_{1}$ given above is not very accurate almost uniquely 
because dimensionless magnetization $m > h /2\hbar (D+E)$ for any 
reasonable $S$ because of admixture into excited states.  For example 
with $E/D = 0.35$ and $S=10$ it was found that $m = 1.0353718 (h 
/2\hbar (D+E))$ was indistinguishable from the numerical data for $m$ 
in the range $0< h < h_{1}$ and gave with five figure precision the 
numerical $m_{1}$.  (Notice $ m_{1}\hbar h = 4E$ while $(\hbar 
h/m_{1}) = 2(D+E)$.)

It is also possible to argue for the local field $\sqrt{(D+E)/2E} h$ 
in terms of the two level model.  For an integer spin system for the 
ground state with $h=h_{\ell} =0$ this is ${\cal H} = 2 S h_{\ell} 
S_{z} + \Delta_{0} S_{x}$ where, as above, $ \Delta_{0}$ is the tunnel 
splitting.  For half-integer spin, following Kramer's theorem this 
slitting is absent.  However since the transverse field breaks time 
reversal symmetry it should create a splitting.  The only satisfactory 
fashion of introducing this transverse Zeeman effect without 
introducing a new parameter is to write ${\cal H} = 2 S h_{\ell} S_{z} 
+ \Delta_{0} m S_{x}$, where $m$ is the dimensionless magnetic moment 
in the $x$-direction.  Since $m = h/2\hbar (D+E)$ this gives a 
splitting $2(\Delta_{0}/\hbar (D+E))h$.  {\it However\/} from 
Eqn.~(\ref{quatorze}) this is correct {\it only\/} if $h$ is replaced 
by the local applied field $\pi \sqrt{(D+E)/2E} h$ to give a splitting 
$\pi \Delta_{0}(h/2\hbar \sqrt{2E(D+E)})$.  The field $h_{t}$ 
similarly generates a term $ {\pi \Delta } ( h_{t}/\hbar 
\sqrt{2E(D-E)})S_{y}$.  Clearly this model it is subject to 
experimental verification and nicely illustrates the effects of 
squeezing via the different factors involved with the different 
components of the transverse field.

\section{Experiment and spin squeezing}

The lack of symmetry between the $x$ and $y$-directions reflects spin 
{\it squeezing}.  For equal transverse fields $m_{x} / m_{y} = 
\sqrt{(D-E)/(D+E)} = {\sigma_{x}}^{2}/{\sigma_{y}}^{2} = Q$, the 
degree of squeezing.  More directly, e.g., $\sigma_{x}^{2}$ is given 
by summing the intensity of all transitions measured by the 
longitudinal response to an r.f.  field in the $x$-direction.  As 
explained above, the transverse matrix elements within the ground 
doublet are $\sim \Delta_{0} m $, i.e., extremely small.  It follows 
that the transitions with energy $\sim D$ exhausts this sum rule.  
Experimentally neutron scattering which can measure the same 
correlation function in zero static field would seem most appropriate.

\vskip 20pt

\section*{Appendix}

Given that $\hbar^{2} D$ is the dominant anisotropy energy it would 
seem the most logical to use this easy axis for the quantization, 
however much of the physics of the problem corresponds to the fact 
that term proportional to $E$ creates an easy plane.  It is therefore 
useful to re-write $\cal H$ quantized along the hard axis.  To this 
end, first use ${S_{y}}^{2} = \hbar^{2}S(S+1) - {S_{x}}^{2} - 
{S_{z}}^{2}$ whence,
$$
{\cal H} = - (D-E) {S_{z}}^{2} + 2 E {S_{x}}^{2} 
 - E \hbar^{2} S(S+1),
$$
then interchanging the $x$ and $z$ axes results in
$$
{\cal H} = 2 E {S_{z}}^{2} - h S_{z} - (D-E) {S_{x}}^{2} 
- h_{\ell} S_{x}-h_{t}S_{y},
$$
where the constant has been omitted and the fields re-introduced.  

The case when $D<E$ can be transformed to $D^{\prime}>E^{\prime}$ 
where $D^{\prime} = 3E - D$ and $E^{\prime} = D+E$.  To show this, 
first use again ${S_{z}}^{2} = \hbar^{2} S(S+1) - {S_{x}}^{2} - 
{S_{y}}^{2}$ to write:
$$
{\cal H}
= - D S(S+1) 
+ (D+E){S_{x}}^{2} - (E-D){S_{y}}^{2}.
$$
With the interchange of the $z$ and $y$-axes and to within a 
constant:
$$
{\cal H} = 
- (E-D){S_{z}}^{2} + (D+E){S_{x}}^{2} .
$$
Comparing this with ${\cal H} = - (D-E) {S_{z}}^{2} + 2 E {S_{x}}^{2}$
it is seen that,
$$
D+E \Rightarrow 2E, \ \ \ \ 
E- D \Rightarrow D-E
$$
reduces the second to the first version.  These manipulations leave 
the term $hS_{z}$ unchanged.  Hence it must be, in particular, that 
the result for the period $h_{1} = 2 \hbar \sqrt{2E(D+E)}$ be 
invariant under the transformation
$$
E \to {1\over2} \left(D + E \right)\ \ \ \
D \to {1\over2} \left(3 E -  D\right).
$$
Then if $E>D$ it follows that $D^{\prime} = 3E - D> 2E$ while 
$E^{\prime} = D+E < 2E$ so that for the transformed problem 
$D^{\prime} > E^{\prime}$, i.e., only that case when $D>E$ need be 
considered.

Another useful version of $\cal H$ is for $h_{t}=h_{\ell} =0$,
$$
{\cal H} = - (D+E) {S_{x}}^{2} - 2 E {S_{y}}^{2} - h S_{z}.
$$

\end{document}